\documentclass{jps-cp}
\usepackage{txfonts} 

\title{Rapidity Decorrelation from Hydrodynamic Fluctuations and Initial Fluctuations}

\author{Azumi \textsc{Sakai}, Koichi \textsc{Murase} and Tetsufumi \textsc{Hirano}}

\inst{Department of Physics, Sophia University, 7-1 Kioi-cho, Chiyoda, Tokyo 102-8554, Japan}

\email{a-sakai-s4d@eagle.sophia.ac.jp, murase@sophia.ac.jp, hirano@sophia.ac.jp}

\recdate{January 28, 2019}

\abst{Rapidity decorrelation in high energy heavy-ion collisions is one of the hot topics in understanding longitudinal dynamics of the quark gluon plasma (QGP). In this study we employ an integrated dynamical model with full three dimensional relativistic hydrodynamics and perform event-by-event numerical simulations of Pb+Pb collisions at the LHC energy. We analyze factorization ratios to understand rapidity decorrelation from hydrodynamic fluctuations and initial longitudinal fluctuations. We show that  factorization breaking happens due to both hydrodynamic fluctuations and initial longitudinal fluctuations. 
We conclude hydrodynamic fluctuations and initial longitudinal fluctuations are both important in understanding rapidity decorrelation.\\
}

\kword{high-energy nuclear collisions, quark gluon plasma, relativistic fluctuating hydrodynamics, factorization ratios}

\begin{document}
\maketitle

\section{Introduction}
Fluctuations play an important role in understanding dynamics of the quark gluon plasma (QGP) created in relativistic heavy ion collisions. 
For example, anisotropic flow $v_n$ results from event-by-event fluctuations of initial transverse profiles. In this study, we focus on fluctuations in the longitudinal (beam) direction. We include both longitudinal fluctuations of distribution of the QGP in the initial stage and hydrodynamic fluctuations in the intermediate stage.
To investigate the effect of fluctuations, we focus on an phenomenon of the event plane decorrelation. The event plane angle can be obtained at each rapidity. Conventionally, the event plane angle is expected to common along the rapidity in an event. However, the event plane angle can be different along the rapidity since the above fluctuations give randomness to the event plane angle at different rapidity. We analyze this rapidity decorrelation and quantify it by factorization ratio $r_n$. The factorization ratio is measured by some experimental groups although none of the theoretical models has yet completely understood the mechanism of this phenomenon.
We analyze factorization ratio $r_n$ and evaluate rapidity decorrelation from hydrodynamic fluctuations and longitudinal initial fluctuations.

\section{Model}
We employ an integrated dynamical model \cite{model, muraseD, Murase:2016rhl} on an event-by-event basis to describe the whole reaction processes. The integrated dynamical model is composed of three models each of which describes the relevant stage of relativistic heavy ion collisions.

In the initial stage we employ a Monte-Carlo version of the Glauber model to quantify the number density of participants and that of binary colllisions in the transverse plane. The model is extended in the longitudinal direction by using an general purpose event generator PYTHIA. To account for heavy ion collisions, we accumulate inelastic p+p events in PYTHIA by the number of binary collisions estimated above and accept the produced particles using rapidity  $Y$ and transverse momentum $p_{T}$ dependent acceptance function $w(Y, p_{T})$ \cite{Okai, Kawaguchi}
\begin{align}
w(p_T, Y) = w(Y) \times \frac{1}{2} \left[1-\tanh \left(\frac{p_T-p_{T_0}}{\Delta p_T}\right)\right]
+ 1 \times \frac{1}{2} \left[1+\tanh \left(\frac{p_T-p_{T_0}}{\Delta p_T}\right)\right], 
\end{align}
\begin{align}
w(Y) = \frac{Y_b+Y}{2Y_b}\frac{1}{N_A} + \frac{Y_b-Y}{2Y_b}\frac{1}{N_B}.
\end{align}
Here $Y_b$ is beam rapidity, $p_{T_0}$ and $\Delta p_T$ are parameters to divide soft and hard transverse momentum regions.
$N_A$ and $N_B$ are the number of participants in nucleus A and B, respectively.
This parametrization is designed to exhibits that the number of produced hadrons scales with the number of participants in the low $p_{T}$ region and with the number of binary collisions in the high $p_{T}$ region.

We assume the initial entropy density distribution is proportional to the number distribution of particles generated and accepted as above. Thus the initial entropy distribution is
\begin{align}
s(\tau_0, x, y, \eta_s) = \frac{K}{\tau_0} \sum_{i} \frac{1}{\sqrt{2\pi\sigma^2_\eta}}\frac{1}{2\pi\sigma^2_\perp}\exp\left[-\frac{\left(x-x^i\right)^2+\left(y-y^i\right)^2}{2\sigma^2_\perp}-\frac{\left( \eta_s-\eta^i_s\right)^2}{2\sigma^2_\eta}\right].
\end{align}
Here $x$ and $y$ are transverse coordinates, $\eta_s$ is space-time rapidity and $(x^i, y^i, \eta_s^i =Y^i)$ is the position of the $i$ th hadron at the initial time $\tau_{0}$.
We smear the position of produced hadrons with Gaussian function. $\sigma_\perp$ and $\sigma_\eta$ are Gaussian width in the transverse and longitudinal directions, respectively. $K$ is a overall factor to control the absolute value of multiplicity. In this study, we choose $\sigma_\perp = 0.1$ fm, $\sigma_\eta = 0.3$, $\tau_0 = 0.6$ fm and $K = 4.8$  to reproduce centrality dependence of charged particle multiplicity~\cite{dndeta}.

For the hydrodynamic evolution of the QGP, we solve the hydrodynamic equations
in the Milne coordinates with an equation of state, $s95p$-v1.1 \cite{EOS}.
For the fluctuating hydrodynamics, the constitutive equation
for the shear-stress tensor at the second order is written as
\begin{align}
\label{eq:shear_stress_tensor}
\tau_\pi{\Delta^{\mu\nu}}_{\alpha\beta}u^{\lambda}\partial_{\lambda}\pi^{\alpha\beta}+\left(1+\frac{4}{3}\tau_\pi\partial_{\lambda}u^\lambda \right) \pi^{\mu\nu} &= 2\eta{\Delta^{\mu\nu}}_{\alpha\beta}\partial^\alpha u^\beta+\xi^{\mu\nu}.
\end{align}
Here $\eta$ is shear viscosity, $u^{\mu}$ is four fluid velocity and 
${\Delta^{\mu\nu}}_{\alpha\beta}=\frac{1}{2}\left({\Delta^\mu}_\alpha{\Delta^\nu}_\beta+{\Delta^\mu}_\beta{\Delta^\nu}_\alpha \right)-\frac{1}{3}{\Delta^{\mu\nu}}\Delta_{\alpha\beta}$. Relaxation time $\tau_{\pi}$ is included so that the system obeys causality.
We set $\eta/s=1/4\pi$ for shear viscosity and $\tau_\pi=3/4\pi T$ for relaxation time~\cite{relaxationtime1, relaxationtime2}.
To include the fluctuation term $\xi^{\mu \nu}$, we consider fluctuation-dissipation relations
\begin{gather}
\label{eq:FD}
\langle\xi^{\mu\nu}(x)\xi^{\alpha\beta}(x')\rangle = 4\eta T\Delta^{\mu\nu\alpha\beta}\delta_\lambda^{(4)}(x-x').
\end{gather}
Here we smear the delta function with a cutoff parameter $\lambda$ in actual simulations.
After macroscopic hydrodynamic simulations, we switch description to microscopic kinetic theory. Here we use Cooper-Frye formula~\cite{CooperFrye} for particlization and use a hadron cascade model JAM~\cite{Nara:1999dz} for description of space-time evolution of hadrons. For switching temperature we use $T_{\mathrm{sw}}=155$ MeV.
\section{Result}

In the following, we show the results for Pb+Pb  collisions at $\sqrt{s_{NN}}=2.76$ TeV.
We analyze factorization ratio for the $n$ th order of harmonics
\begin{align}
\label{eq:factorization_ratio}
r_n(\eta^a_p,\eta^b_p) &= \frac{V_{n\Delta}(-\eta^a_p,\eta^b_p)}{V_{n\Delta}(\eta^a_p,\eta^b_p)},&  V_{n\Delta} &= \langle{\cos (n\Delta\phi)} \rangle.
\end{align}
Here $\Delta\phi$ is a difference of azimuthal angle between two particles in the separated rapidity regions.
When $r_n(\eta^a_p,\eta^b_p)=1$, there is a unique event plane for each event. While, when $r_n(\eta^a_p,\eta^b_p)<1$,  the event plane angle depends on rapidity which brings factorization breaking and rapidity decorrelation.
We analyze $r_n(\eta^a_p,\eta^b_p)$ with fluctuating hydrodynamics and viscous hydrodynamics which has no fluctuating terms ($\xi^{\mu \nu}$=0 in Eq.~(\ref{eq:shear_stress_tensor})) in the hydrodynamic stage. Note that both models include initial longitudinal fluctuations, which is the first attempt within this framework.
\begin{figure}[tbh]
\includegraphics[bb=0 10 250 210]{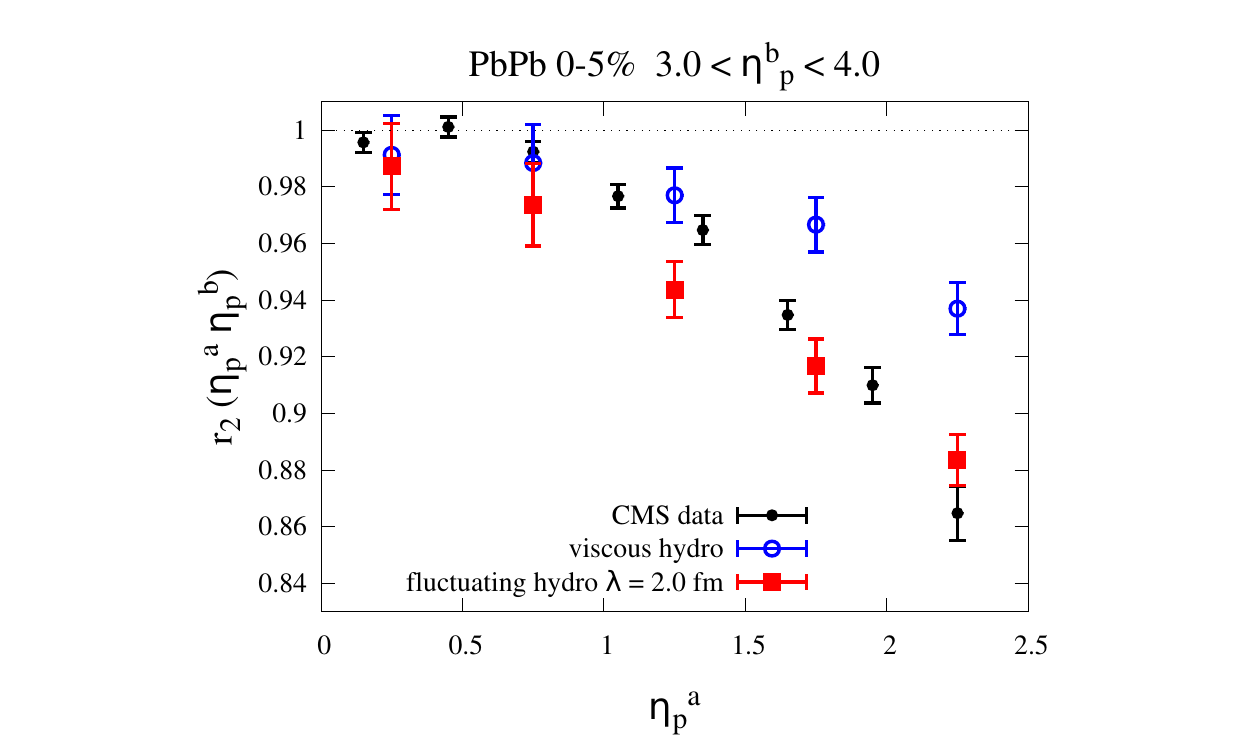}\\
\includegraphics[bb=0 10 250 210]{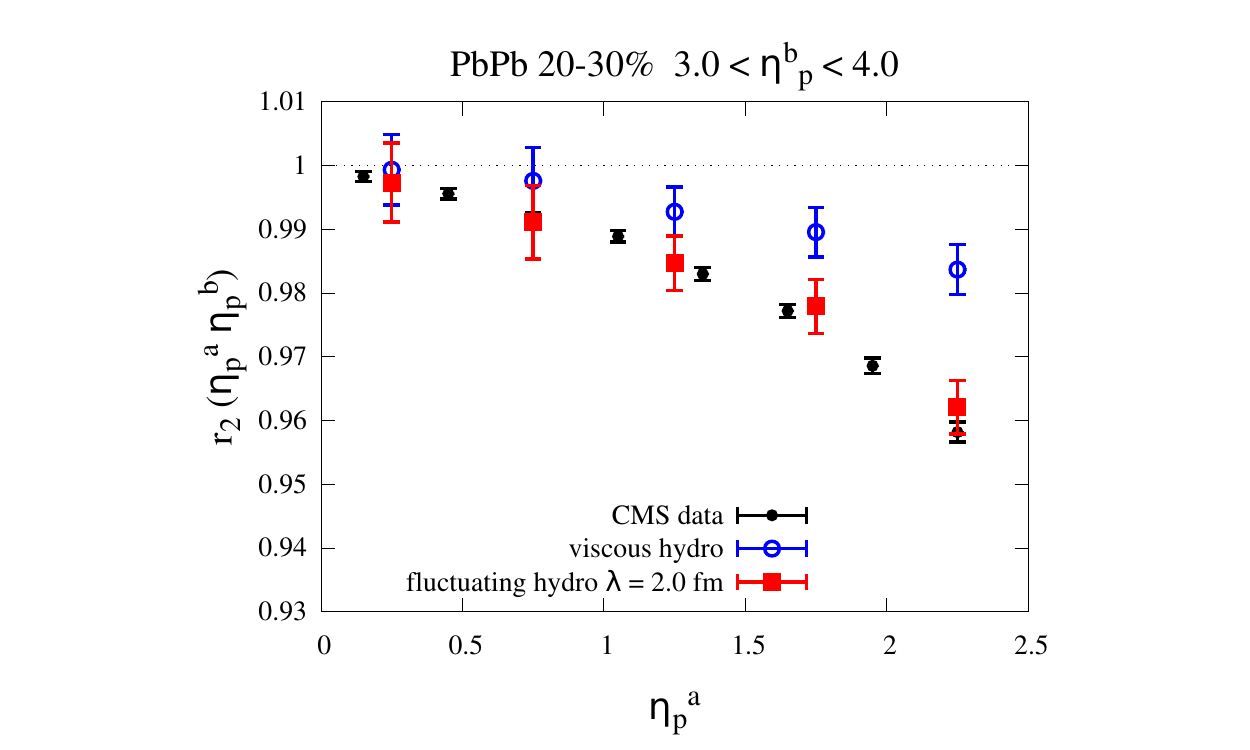}\\
\includegraphics[bb=0 10 250 210]{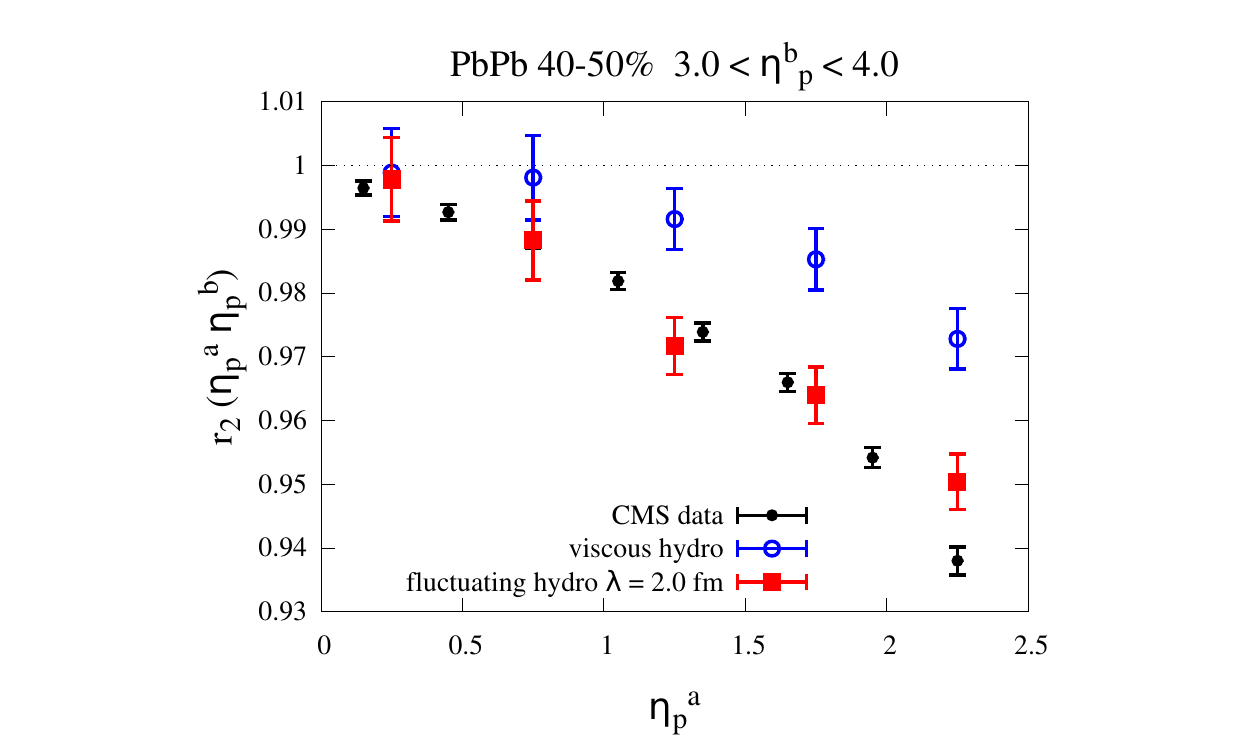}
\caption{$\eta^a_p$ dependence of $r_2(\eta^a_p, \eta^b_p)$ with reference rapidity $3.0 < \eta^b_p < 4.0$. From top to bottom, results in Pb+Pb collisions at $\sqrt{s_{NN}}=2.76$ TeV
are shown for centrality  0-5, 20-30 and 40-50\%. Results from fluctuating hydrodynamics with initial longitudinal fluctuations (red square), viscous hydrodynamics with initial longitudinal fluctuations (blue circle) and CMS data (black diamond)~\cite{rn:CMSdata} are shown.}
\label{f3}
\end{figure}

We generate 4000 hydrodynamic events 
for each of which we perform 100 times hadronic cascade simulations.
Figure \ref{f3} shows the result of $\eta^a_p$ dependence of $r_2(\eta^a_p, \eta^b_p)$
in Pb+Pb collisions at $\sqrt{s_{NN}}=2.76$ TeV. 
One sees $r_{2}$ from viscous hydrodynamics with initial longitudinal fluctuations decreases with $\eta^a_p$. This is in contrast to the result $r_{2} \sim 1$ obtained in our previous study without initial fluctuations \cite{Sakai}, and indicates the importance of the initial longitudinal fluctuations of the entropy density distribution to the factorization breaking as well.
While, $r_{2}$  from fluctuating hydrodynamics with initial longitudinal fluctuations decreases with $\eta^a_p$ more significantly than that from viscous hydrodynamics does. 
It turns out the results from fluctuating hydrodynamics with initial longitudinal fluctuations are closer to experimental data~\cite{rn:CMSdata}. Therefore, we conclude hydrodynamic fluctuations are as important as initial longitudinal fluctuations of profiles in understanding rapidity decorrelation.

\section{Summary}
We simulated relativistic heavy ion collisions using an integrated dynamical model based on full three-dimensional fluctuating hydrodynamics and viscous hydrodynamics with longitudinal fluctuations of initial profiles.
This is the first analysis of rapidity decorrelation from a model including both hydrodynamic fluctuations and longitudinal fluctuations of initial profiles at the same time.
We examined the effect of both fluctuations by analyzing factorization ratio $r_2(\eta^a_p, \eta^b_p)$.
From a fact that  factorization ratios $r_2(\eta^a_p, \eta^b_p)$ decreases with $\eta^a_p$ in viscous hydrodynamic model,  event plane angles decorrelate along rapidity due to longitudinal fluctuations of initial profiles.
In the fluctuating hydrodynamics model with initial longitudinal fluctuations, factorization ratios $r_2(\eta^a_p, \eta^b_p)$ decrease more. Therefore, even more rapidity decorrelation occurs due to hydrodynamic fluctuations. Since the result with both  hydrodynamic fluctuations and longitudinal fluctuations of initial profiles is closer to experimental data, we conclude the effects of these fluctuations are both important in understanding rapidity decorrelation.
\section*{Acknowledgment}
This work was supported by JSPS KAKENHI Grant Numbers JP18J22227(A.S.) and JP17H02900(T.H.).

\end{document}